\documentclass[twocolumn,epjc3]{svjour3} 
\usepackage{amssymb}
\usepackage{amsmath}
\usepackage{comment}
\usepackage{lineno}

\RequirePackage{graphicx,subfig,multirow,ulem}

\DeclareGraphicsExtensions{.pdf}
\RequirePackage{mathptmx} 
\RequirePackage[numbers,sort&compress]{natbib}
\RequirePackage[colorlinks,citecolor=blue,urlcolor=blue,linkcolor=blue]{hyperref}

\journalname{Eur. Phys. J. C}

\newcommand{\DBD}{0$\nu\beta\beta$}
\newcommand{\PO}{$^{210}$Po}
\newcommand{\KF}{$^{40}$K}
\newcommand{\TEO}{$\mathrm{TeO}_2$}
\newcommand{\TEHT}{$^{130}\mathrm{Te}$}

\newcommand{\TL}{$^{208}\mathrm{Tl}$}
\newcommand{\THO}{$^{232}\mathrm{Th}$}

\newcommand{\CUORICINO}{Cuoricino}
\newcommand{\CUORE}{CUORE}

\newcommand{\MC}{Monte Carlo}
\newcommand{\ckky}{\un{counts/(keV\,kg\,y)}}

\newcommand{\CERE}{Cherenkov}
\providecommand*{\un}[1]{\ensuremath{\mathrm{~#1}}}

\begin{document}       

\title{TeO$_2$ bolometers with \CERE\ signal tagging: towards next-generation neutrinoless double beta decay experiments}

\author{
{N.~Casali}\thanksref{INFN-LNGS,UNIAQ}
\and
{M.~Vignati}\thanksref{UNISAP,INFN-RM1,e1}
\and
{J.W.~Beeman}\thanksref{LBNL}
\and
{F.Bellini}\thanksref{INFN-RM1,UNISAP}
\and
{L.~Cardani}\thanksref{INFN-RM1,UNISAP}
\and
{I.~Dafinei}\thanksref{INFN-RM1}
\and
{S.~Di~Domizio}\thanksref{INFN-GE,UNIGE}
\and
{F.~Ferroni}\thanksref{INFN-RM1,UNISAP}
\and
{L.~Gironi}\thanksref{INFN-MIB,UNIBIC}
\and
{S.~Nagorny}\thanksref{INFN-GSSI}
\and
{F.~Orio}\thanksref{INFN-RM1}
\and
{L.~Pattavina}\thanksref{INFN-LNGS}
\and
{G.~Pessina}\thanksref{INFN-MIB}
\and
{G.~Piperno}\thanksref{INFN-RM1,UNISAP}
\and
{S.~Pirro}\thanksref{INFN-LNGS}
\and
{C.~Rusconi}\thanksref{INFN-MIB}
\and
{K.~Sch\"affner}\thanksref{INFN-LNGS}
\and
{C.~Tomei}\thanksref{INFN-RM1}
}
\thankstext{e1}{e-mail: marco.vignati@roma1.infn.it}

\institute{
{INFN  Laboratori Nazionali del Gran Sasso, I-67010 Assergi (AQ) - Italy}\label{INFN-LNGS}
\and
{Dipartimento di Scienze Fisiche e Chimiche, Universit\`{a} degli studi dell'Aquila, I-67100 Coppito (AQ) - Italy}\label{UNIAQ}
\and
{Dipartimento di Fisica, Sapienza Universit\`{a} di Roma, I-00185 Roma - Italy}\label{UNISAP}
\and
{INFN Sezione di Roma, P.le~A.~Moro~2, Roma I-00185, Italy}\label{INFN-RM1}
\and
{Lawrence Berkeley National Laboratory , Berkeley, California 94720, USA}\label{LBNL}
\and
{INFN  Sezione di Genova, I-16146 Genova - Italy}\label{INFN-GE}
\and
{Dipartimento di Fisica, Universit\`{a} di Genova, I-16146 Genova - Italy}\label{UNIGE}
\and
{INFN  Sezione di Milano Bicocca, I-20126 Milano - Italy}\label{INFN-MIB}
\and
{Dipartimento di Fisica, Universit\`{a} di Milano Bicocca, I-20126 Milano - Italy}\label{UNIBIC}
\and
{INFN Gran Sasso Science Institute, I-67100 L`Aquila - Italy}\label{INFN-GSSI}
}

\maketitle

\begin{abstract}
CUORE, an array of 988 TeO$_2$ bolometers, is about to
be one of the most sensitive experiments searching for neutrinoless
double-beta decay. Its sensitivity could be further improved by removing
the background from $\alpha$ radioactivity. A few years ago it was pointed out that the signal from $\beta$s can be tagged by detecting the emitted \CERE\
light, which is not produced by $\alpha$s.
In this paper we confirm this possibility. For the first time we measured
the \CERE\ light emitted by a CUORE crystal, and found it to be
100~eV at the $Q$-value of the decay. To completely reject the $\alpha$ background,
we compute that one needs light detectors with baseline noise below 20~eV~RMS, a value
which is 3-4 times smaller than the average noise of the bolometric
light detectors we are using.  We point out that an improved light detector
technology must be developed to obtain TeO$_2$ bolometric experiments
able to probe the inverted hierarchy of neutrino masses.
\end{abstract}

\PACS{23.40.-s, 07.57.Kp, 29.40.Ka}
\keywords{Neutrinoless double beta decay, bolometer, \CERE\ detector}


\section{Introduction}
Neutrinoless double beta decay (\DBD) is a  process that violates the lepton number conservation law by two units, in which
a parent nucleus decays into a daughter nucleus and emits two $\beta$ particles.
Unlike the process accompanied by the emission of two neutrinos, allowed by the Standard Model and observed in several nuclei,
\DBD\ has not yet been observed. 
Its discovery would reveal physics beyond the Standard Model: it would tell us that neutrinos, unlike all other elementary fermions,  are Majorana particles,
and would point to Leptogenesis as the origin of the matter-antimatter asymmetry after the Big Bang (for a recent review see for example \cite{Bilenky:2012qi} and references therein). The experimental signature is very clear, a peak in the sum energy spectrum of the $\beta$s at the $Q$-value of the decay.

Bolometers proved to be good detectors to search for \DBD, thanks to the high variety of isotopes that can be studied,
the excellent energy resolution, and the low background they can achieve.
The \CUORE\ experiment~\cite{ACryo,Artusa:2014lgv} will search for the \DBD\ of \TEHT\ with an array of 988~\TEO\ bolometers,
cryogenic calorimeters working at a temperature around 10\un{mK}. 
Each bolometer weighs 750\un{g}, for a total active mass of 741\un{kg},  206\un{kg} of which are \TEHT\ (34.2\% natural abundance~\cite{Fehr200483} in tellurium). The energy resolution at the $Q$-value of the decay, 2528\un{keV}~\cite{Redshaw:2009zz}, is expected to be 5\un{keV~FWHM}. 
\CUORE\ is under construction at Laboratory Nazionali del Gran Sasso (LNGS) in Italy, and will start to take data in 2015.
 
The technology of \TEO\ bolometers has been demonstrated by \CUORICINO, a 40\un{kg} tower of 62 bolometers that,  with $19.75\un{kg\,y}$ of \TEHT\ data, set a lower limit to the decay half-life of $2.4\cdot 10^{24}\un{y}$ at  90\% C.L.~\cite{Andreotti:2010vj}. The analysis of the data pointed out that the main source of background in the energy region of interest (ROI) for the \DBD\ consisted in $\alpha$ particles
generated by natural radioactivity of the copper structure holding the crystals. To reduce it, the \CUORE\ collaboration developed techniques to clean the copper and to assemble the detector in ultra radiopure environments. The success of this effort has been recently demonstrated by the CUORE-0 experiment, an array of 52 bolometers that reached an $\alpha$ background index of $0.019\pm 0.002\ckky$, a factor 6 less than \CUORICINO~\cite{Aguirre:2014lua}. 
The background in \CUORE, however, is still foreseen to be dominated by $\alpha$ particles, limiting the sensitivity to the \DBD\ half-life to around $10^{26}$ years in 5 years of data taking. This corresponds to an effective neutrino Majorana mass that ranges, depending on the choice of the nuclear matrix element, from $40$ to $100\un{meV}$,  values that are quite far from covering the entire interval of masses corresponding to the inverted hierarchy scenario, that ranges from $10$ to $50\un{meV}$~\cite{Bilenky:2012qi}.

The background can be reduced by detecting the small amount of  \CERE\ light that is emitted by interacting particles in \TEO\ crystals. In fact, at the energy scale of interest for \DBD, the $\beta$s (signal) are above threshold for \CERE\ emission, while $\alpha$ particles (background) are not~\cite{TabarellideFatis:2009zz}. 
In a previous paper~\cite{Beeman:2011yc} we operated a 117\un{g} \TEO\ bolometer surrounded by a 3M VM2002 reflecting foil, monitoring a crystal face with a germanium bolometer acting as light detector. In coincidence with the heat released in the \TEO\ we were able to detect the light emitted by $\beta/\gamma$ particles, which amounted to 173\un{eV} at 2528\un{keV}. The crystal was doped with natural samarium, which contains $^{147}$Sm, an $\alpha$-unstable isotope  with $Q=2310\un{keV}$. The light detected from these decays was compatible with zero,  confirming that at the \DBD\ energy scale no light is emitted by $\alpha$s.  Finally, room temperature tests confirmed that the light emitted by particles interacting in \TEO\ can be ascribed to the sole \CERE\ emission, excluding a contribution from the scintillation~\cite{Casali:2013bva}.

In this paper we present the results of a test conducted on a CUORE bolometer, i.e. a 750\un{g} crystal, 6 times larger than that used in our previous work and without samarium doping. The results confirm that the $\alpha$ discrimination in \CUORE\ is possible,
but the light signal is small and requires light detectors with higher sensitivity than that provided by bolometers.

\section{Experimental setup}
\begin{figure}[tb]
\centering
\includegraphics[width=0.48\textwidth]{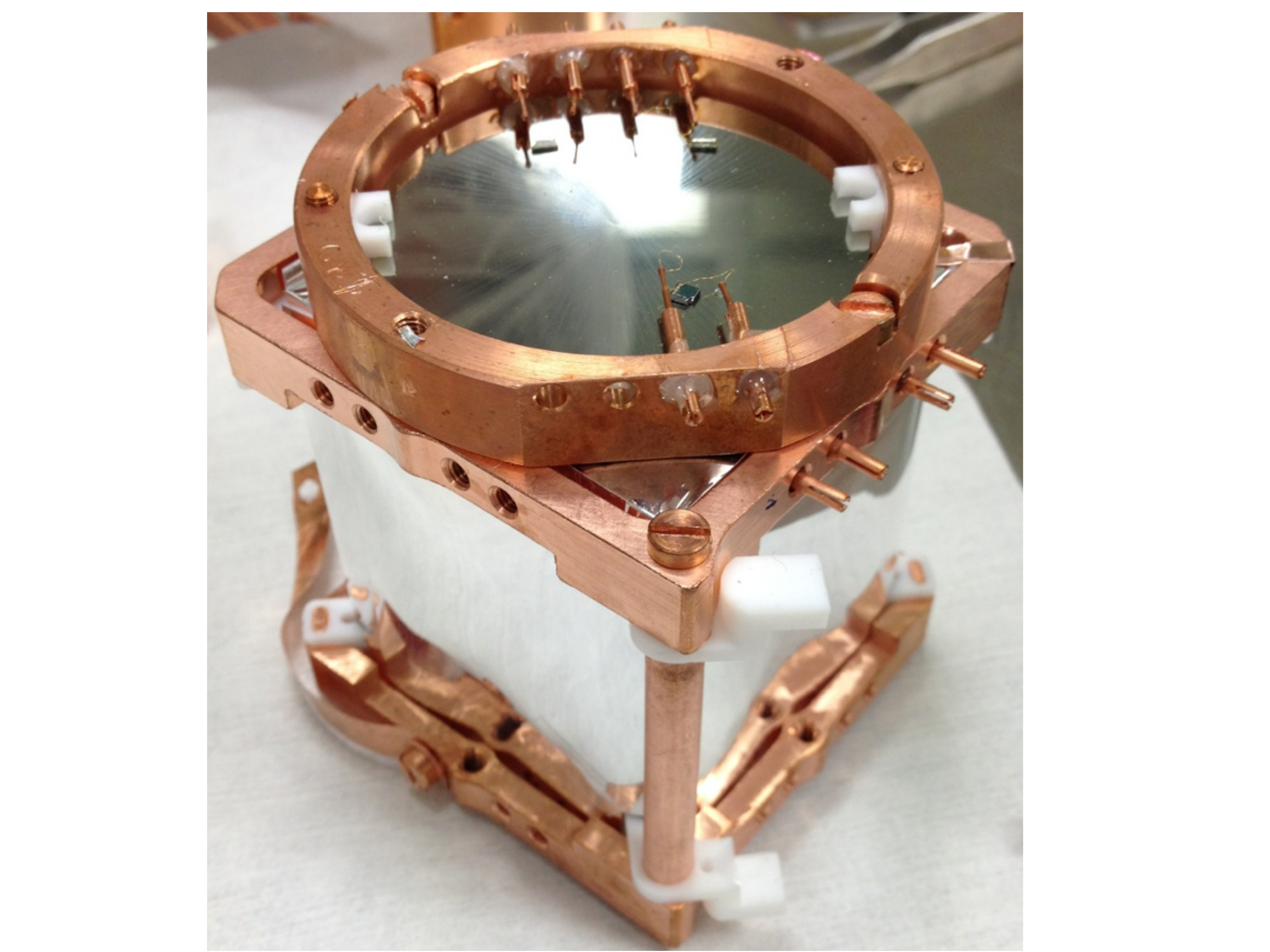}
\caption{The \TEO\ crystal in the copper holder, surrounded by a 3M VM2002 light reflector and monitored by the germanium bolometric light detector.}
\label{fig:setup}
\end{figure}

The \TEO\ crystal comes from samples of the \CUORE\  batches
used to check the radiopurity and the bolometric performances during the production~\cite{Alessandria:2011vj},
and therefore is identical to the crystals that are currently being mounted in \CUORE.
The crystal is a $5\times5\times5\un{cm^3}$ cube with translucent faces, two opposite of which 
have a better polishing quality, close to optical polishing grade. 
All faces are surrounded by the VM2002 light reflector except for an optical one that is monitored by a 5\un{cm} in diameter, $300\un{\mu m}$ thick germanium light detector (LD)~\cite{Beeman:2013zva} (Fig.~\ref{fig:setup}).
Both the \TEO\ crystal and the germanium are operated as bolometers, using a neutron transmutation doped germanium (NTD-Ge) thermistor as temperature sensor~\cite{Itoh}. The detectors are held in a copper structure by means of teflon (PTFE) supports, anchored to the mixing chamber of a dilution refrigerator.
The setup is operated in the CUORE/LUCIFER R\&D cryostat, in the Hall C of LNGS~\cite{Pirro:2006mu}.

As in Ref.~\cite{Beeman:2011yc}, the read-out of the thermistor is performed using the \CUORICINO\ electronics~\cite{Arnaboldi:2004jj}. The analog signals are filtered by 6-pole active Bessel filters~\cite{Arnaboldi:2010zz} and then fed into an 18-bit National Instrument PXI analog-to-digital converter (ADC), the same system being used in CUORE-0. The filter cutoff and the ADC sampling frequency are set to 12\un{Hz} and 125\un{Hz} for the \TEO, respectively, and  to 120\un{Hz} and 2000\un{Hz} for the LD, respectively.  The trigger is software generated on each bolometer. When it fires, waveforms 5\un{s} long on the \TEO\ and 250\un{ms} long on the LD are saved on disk. Additionally, when the trigger fires on the \TEO, the waveform on the LD is acquired irrespective of its own trigger. 

To maximize the signal to noise ratio, the waveforms are processed offline with the optimum filter algorithm~\cite{Radeka:1966,Gatti:1986cw}. On the \TEO\ the pulse  is identified with a peak finder algorithm, and the amplitude is evaluated as the maximum of the peak.
On the LD, to eliminate noise artifacts at the threshold, the pulse amplitude is evaluated at the characteristic time delay of the LD response with respect to the pulse on the \TEO, which is estimated in calibration runs using events generated by particles interacting in both detectors (for more details see Ref.~\cite{Piperno:2011fp}).

The light detector is exposed to a permanent $^{55}$Fe source, providing 5.9 and 6.5\un{keV} calibration X-rays.
The typical rise and decay times of the  pulses are 2.6 and 6\un{ms}, respectively, while the energy resolution at the iron peaks and at the baseline is 135 and 72\un{eV~RMS}, respectively. 
To calibrate the \TEO\ and to generate events in the \DBD\ region, the setup is illuminated by a \THO\ source placed outside the cryostat. The rise and decay times of the \TEO\ pulses are 40 and 532\un{ms}, respectively,  values that are similar to the CUORE-0 ones~\cite{Aguirre:2014lua}. 

The energy resolution at the 2615~keV $^{208}$Tl peak from the thorium source is 11.5\un{keV~FWHM}, worse than the 5.7\un{keV} FWHM obtained averaging all the CUORE-0 bolometers. This might be due to the different working temperature, which was chosen higher than in CUORE-0 (20~mK instead of 10~mK) in order to improve the energy resolution of the light detector (see  Ref.~\cite{Beeman:2013zva} for details). The worse energy resolution of the \TEO\  bolometer does not affect our results, since the attention is focused on the light signal.

\section{Results}\label{sec:results}
\begin{figure}[tb]
\centering
\includegraphics[width=0.48\textwidth]{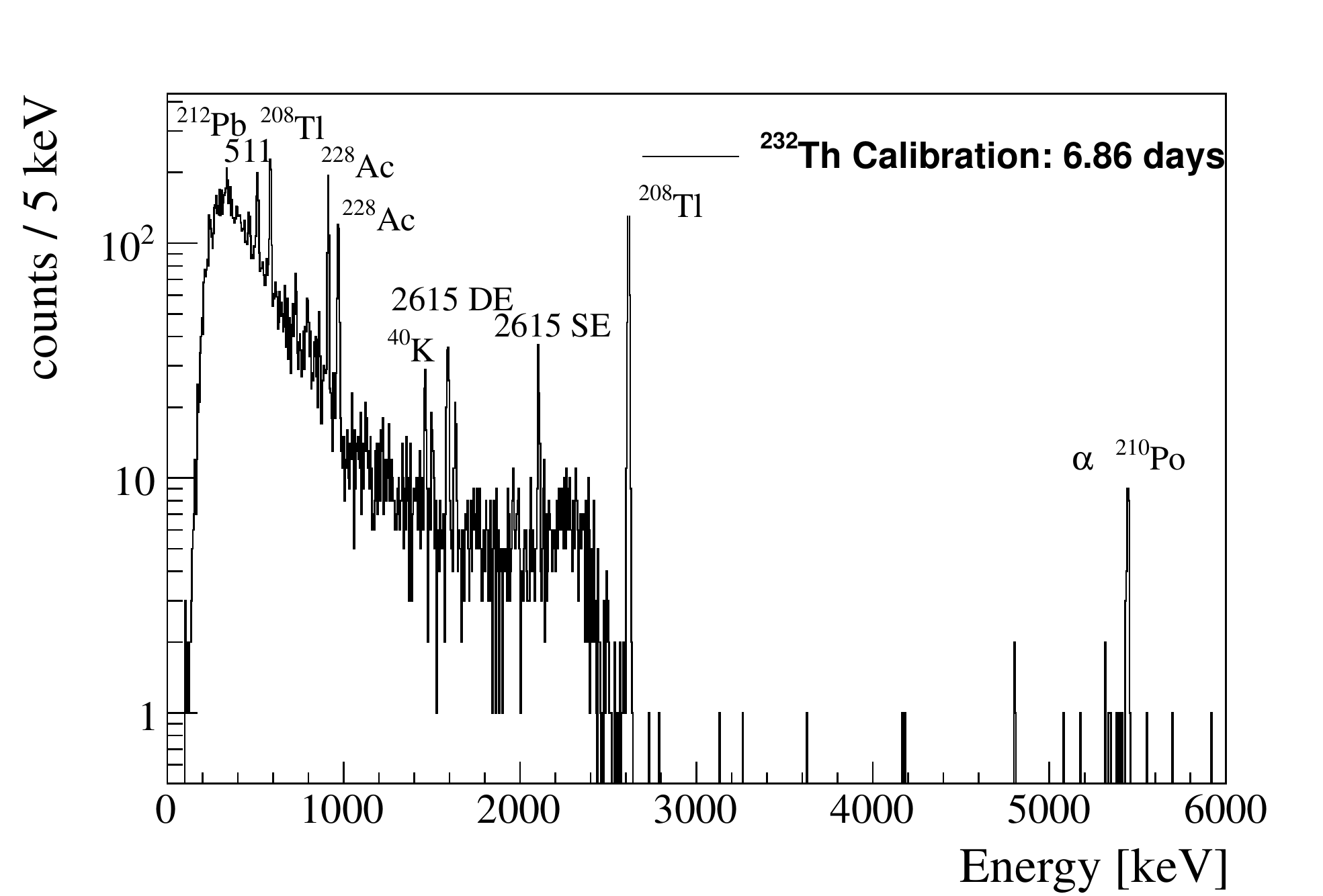}
\caption{Energy spectrum acquired by the \TEO\ crystal. All the labeled peaks are $\gamma$s,  except for the single and double escape peaks of the 2615\un{keV} $\gamma$ from \TL, which are $e^- + e^+ +\gamma$ and $e^- + e^+$ events, respectively, and for the events
around 5.4\un{MeV},  which are generated  by the  $\alpha$-decay of \PO\ in the crystal.}
\label{fig:energyspectrum}
\end{figure}

The energy spectrum acquired from the \TEO\ bolometer in 6.86 days of data taking is shown in Fig.~\ref{fig:energyspectrum}.
The peak around 5400\un{keV} is due to the $\alpha$-decay of \PO, a natural contamination of the \TEO\ crystal observed also in the 117\un{g} detector and in CUORE-0. The remaining peaks are $\gamma$s from the \THO\ source, except for the peak at 1461\un{keV}, which is a $\gamma$ from \KF\ contamination of the cryostat. Both the single (SE) and double escape (DE) peaks of the 2615 keV $\gamma$ from \TL\ are visible. The presence of the DE peak is of particular interest because it is a single site production of a $e^-$ and of a $e^+$, a process similar to the \DBD. 
%

\begin{figure}[tb]
\centering
\includegraphics[width=0.48\textwidth]{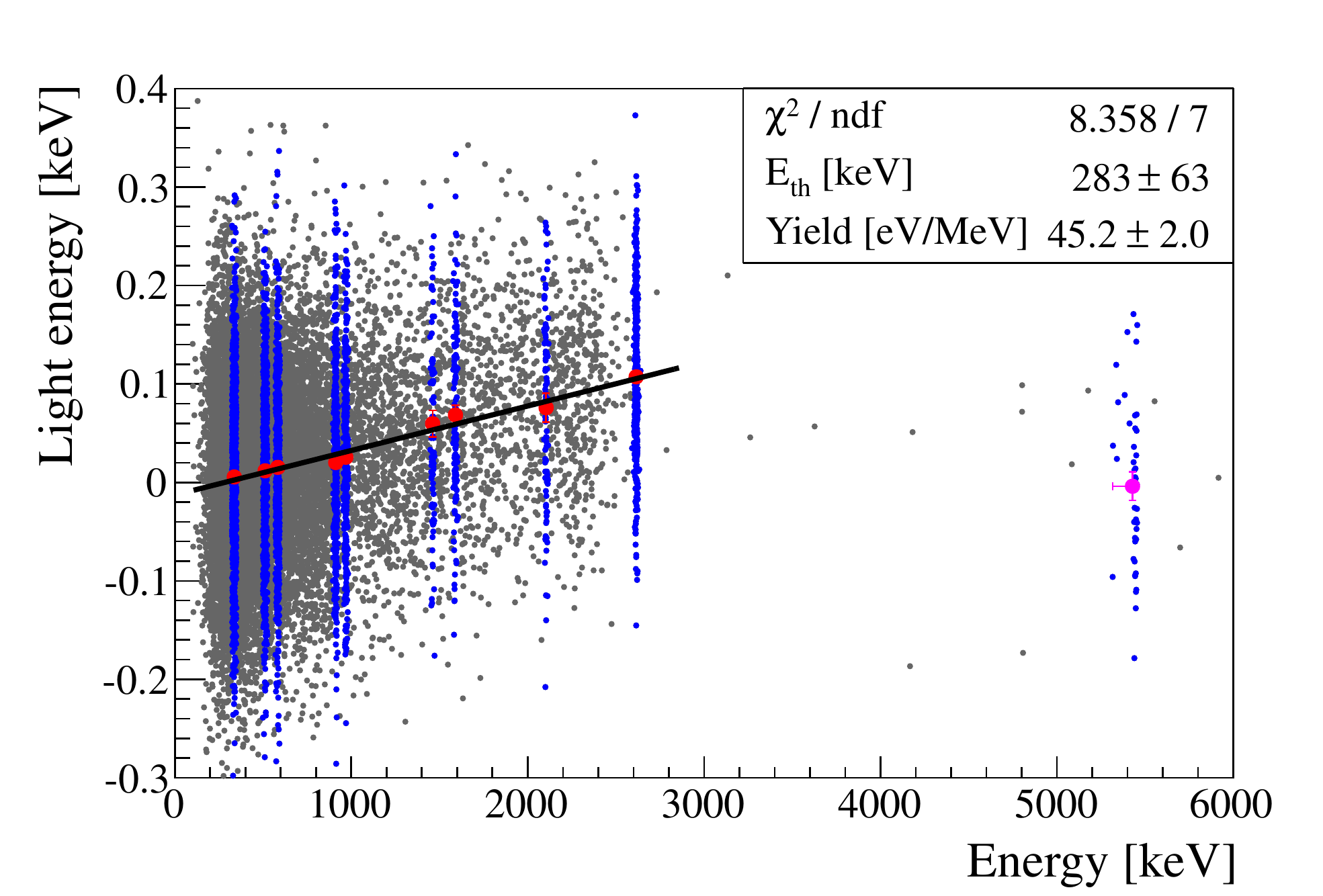}
\caption{(Color online) Detected light versus calibrated heat in the \TEO\ bolometer for all the acquired events (gray) and for the events belonging to the peaks labeled in Fig~\protect\ref{fig:energyspectrum} (blue).  The mean light is clearly energy dependent for the $\gamma$ peaks (red circles below 3 MeV) and compatible with zero for the $\alpha$-decay of the \PO\ (pink circle at 5.4 MeV).  }
\label{fig:lightvsheat}
\end{figure}

The light detected versus calibrated heat in the \TEO\ crystal is shown in Fig.~\ref{fig:lightvsheat}.
The distribution of the light corresponding to each peak in Fig.~\ref{fig:energyspectrum} (blue dots in the figure) is fitted with a Gaussian, the mean of which is overlaid onto the figure. The mean light from the $\alpha$-decay of \PO\  is found to be $<L_\alpha> = -3.9\pm 14.5\un{eV}$, i.e. compatible with zero. The mean light from the $\gamma$ peaks is fitted with a line   $<L_{\beta/\gamma}> = {\rm LY}\cdot ({\rm Energy-E_{th}})$, with ${\rm E_{th}} = 280\pm 60\un{keV}$ and ${\rm LY} = 45\pm 2\un{eV/MeV}$. 
The standard deviations of the light distributions are found compatible with the baseline noise of the LD, which therefore appears as the dominant source of fluctuation, hiding any possible dependence on the position of the interaction in the \TEO\ crystal or statistical fluctuations of the number of photons.  
As in our previous work, the light from the DE peak is compatible with the light from $\gamma$s, indicating that the fitted line can be used to predict the amount of light detectable from \DBD\ events. We compute  $101.4\pm 3.4$\un{eV} of light for a $\beta/\gamma$ event with \DBD\ energy, 72\un{eV} less than the light detected at the same energy in the 117\un{g} detector. 

The detected light at the \DBD\ is small, at the same level of the LD noise, and does not allow one to perform an event by event rejection of the $\alpha$ background. 
As indicated in Ref.~\cite{TabarellideFatis:2009zz}, the emitted \CERE\ light amounts to several hundreds of eV, a much higher value than what we detect.

To increase the light collection efficiency, we applied different modifications to the setup:
1) we changed the VM 2002 light reflector to aluminum foils. Aluminum is expected to have higher reflectivity in the UV band, the region where the \CERE\ emission is more intense.  Nevertheless, the amount of light detected is 25\% less than in the case of VM 2002;
2) we removed the VM 2002, which is a specular light reflector, and wrapped the crystal with teflon tape, which is a light diffusor. The amount of light detected is compatible with the VM 2002 measurement;
3) we changed the LD to an identical one, but we coated  the side faced to the \TEO\ with 60\un{nm} of SiO$_2$. It has been demonstrated, in fact, that in the red/infrared band this layer enhances the light absorption by up to 20\%~\cite{Beeman:2012cu,ZnSe2013}. In our application, however, the amount of light detected does not change significantly;
4) we added a second LD, monitoring opposite faces with two different light detectors. The amount of light detected from each LD is found to be the 50\% of the amount detected with a single LD. This causes an overall decrease of the signal to noise ratio, because each LD adds its own noise;
5) we replaced the crystal with a cylindrical one, 4\un{cm} in diameter and in height. Again the amount of light detected does not change.

Summarizing, none of the above trials succeeded in providing a significant increase of the light collection efficiency, indicating that most of the light is absorbed by the \TEO\ crystal. This is due to the high refractive index of the \TEO\ crystal ($n\sim 2.4$~\cite{PhysRevB.4.3736}): many photons are  reflected internally several times up to absorption. This effect is confirmed by the higher light yield obtained with the small (117~g) crystal and by preliminary results from simulations of the light collection.

The setup providing the highest light signal, around 100\un{eV} at the \DBD, consists in a single LD with the crystal surrounded by the VM 2002 reflector or wrapped with teflon tape. 

\section{Perspectives}

 The recent CUORE-0 result restricted the prediction of the amount of $\alpha$ background in \CUORE\ from $B_\alpha = 0.01-0.04$ to $B_\alpha=0.01\ckky$~\cite{Aguirre:2014lua}, while the ultimate source of background, due to $\beta/\gamma$ radioactivity from  the setup, still amounts to $B_{\beta/\gamma} = 0.001\ckky$. From these numbers and from the specs of \CUORE\ we perform toy \MC\ simulations to estimate the 90\%\un{C.L.}  sensitivity of \CUORE\ equipped with light detectors.

\begin{figure}[tb]
\centering
\includegraphics[width=0.48\textwidth]{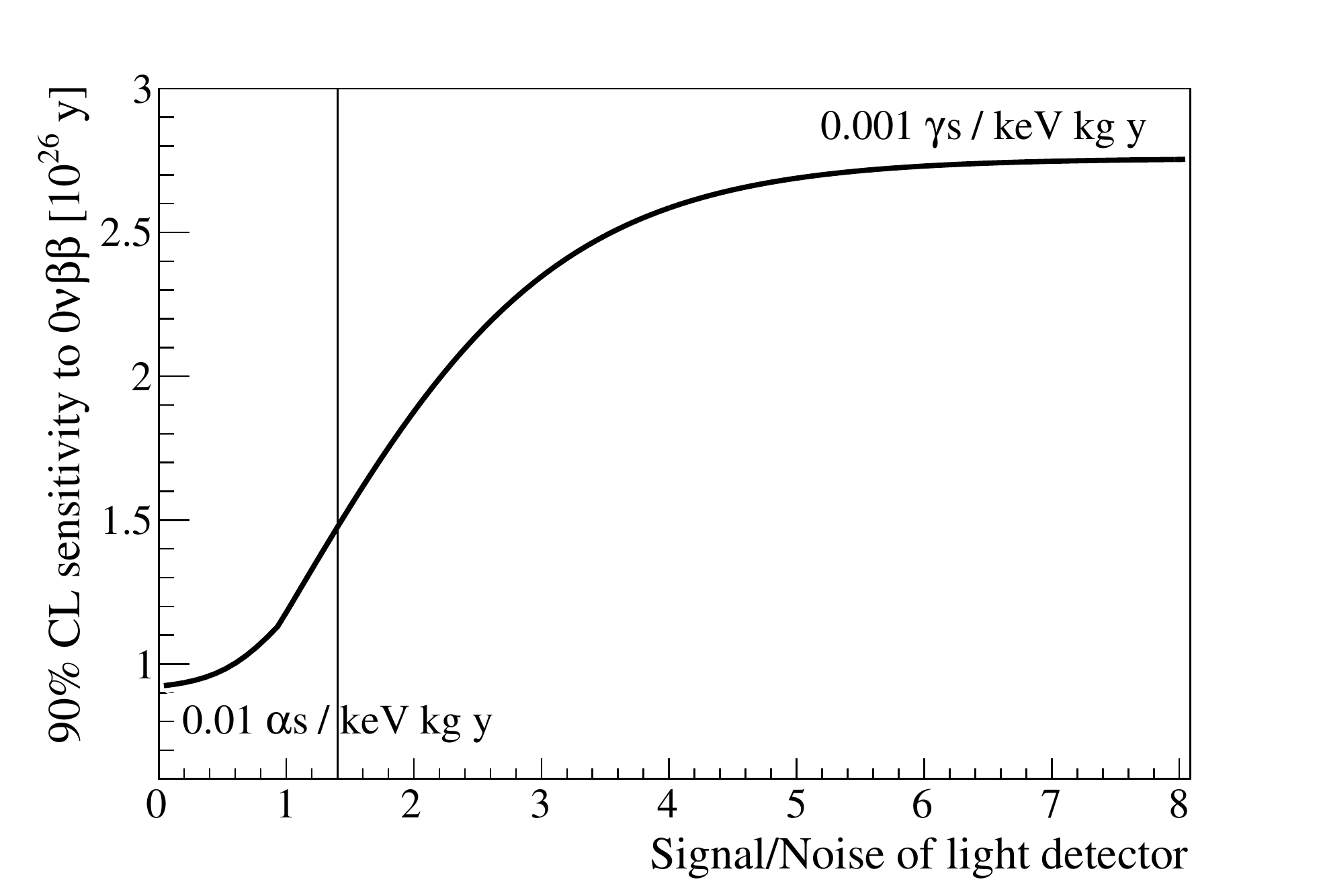}
\caption{90\% C.L. sensitivity to the half-life of \TEHT\ as a function of the signal to noise ratio of the light detectors,
under the reasonable hypothesis of  an $\alpha$ background index in \CUORE\ of $0.01\ckky$. The sensitivity of the experiment without light detectors corresponds to ${\rm S/N = 0}$. When ${\rm S/N > 5}$ the $\alpha$ background is hidden by the
unreducible  background predicted from $\gamma$  interactions, amounting to $0.001\ckky$, and the sensitivity is maximal.  The performance of the light detectors used in this work, ${\rm S/N} = 1.4$, is clearly too low.
}
\label{fig:sensitivity}
\end{figure}

The outcome of a toy experiment is  fitted  in energy with a flat probability density function (pdf) for the background and a Gaussian pdf for the signal, and is simultaneously fitted in light with Gaussians pdfs for the $\alpha$ and $\beta/\gamma$  light distributions. The posterior pdf of the signal events is obtained integrating over the nuisance parameters and assuming a flat prior. The sensitivity  of a single experiment  ($N_{90}$) is computed as the number of signal events corresponding to the 90\% of the posterior cumulative distribution. Several ($\sim 3000$) experiments are generated, and the median of $N_{90}$ is used as estimator of the sensitivity. The entire procedure is repeated while varying the signal to noise ratio in the light detector (Fig.~\ref{fig:sensitivity}).

From the figure  one sees that the application of light detectors to \CUORE\ would increase its 90\% C.L. sensitivity to the half-life of \TEHT\ to  $2.7\cdot 10^{26}\un{y}$, a factor 3 higher than \CUORE\ without light detectors.
To achieve this goal one needs a signal to noise ratio in the light detector greater than 5, a value that is far from that featured by the setup in this work, equal to $101/72=1.4\un{eV/eV}$.

From the results presented the increase of the light signal is difficult, and therefore  to upgrade \CUORE\
 light detectors able to provide a noise level below $20\un{eV~RMS}$ are needed. Other than trying to improve the NTD technology, there are at least two possible alternatives. The use of phonon-mediated transition edge sensors (TES), as in the CRESST dark matter experiment~\cite{Angloher:2011uu}, or the use of phonon-mediated kinetic inductance detectors (KID), as recently proposed in Ref.~\cite{CalderLTD13}.
The TES technology has already proved to reach very good noise levels, but the implementation of 988 light detectors
implies a  complicated readout, mainly because of the cryogenic SQUID amplifiers that are employed. 
KIDs already proved to be a highly multiplexable technology in astrophysical applications (up to $400$ channels
on the same readout line~\cite{Bourrion:2012td}) but the required energy resolution in our application still needs to be demonstrated. 

\section{Conclusions}
We tested the possibility to discriminate the $\alpha$ background in \CUORE\ by tagging the signal from $\beta$ particles
through the detection of \CERE\ light. The detected light at the \TEHT\ $Q$-value is around 100\un{eV} 
 for $\beta/\gamma$ particles and no light is detected from $\alpha$ interactions, confirming the validity of this technology. However, the signal is small at the same level of the noise of the bolometric light detectors we are using, and does not allow us to perform an event by event discrimination of the background. We tested modifications of the setup, by using different light reflectors or multiple light detectors, but the light yield did not increase. 
 
We are  working on simulations to estimate the fraction of emitted light that escapes the crystal and is eventually
absorbed by the light detector. Critical parameters are the index of refraction and the absorbance of \TEO, which unfortunately are not available in the literature for low temperatures. To this end we are setting up a dedicated measurement. 

Given the results obtained so far, we conclude that, to remove completely the $\alpha$ background in \CUORE,  light detectors with a noise of $20\un{eV~RMS}$ are needed, a factor 3-4 times better than the bolometric light detectors we used in this work. Changing the technology to TES or KID devices could be an alternative, provided that the present readout and sensitivity limits are overcome.

\CUORE\ without $\alpha$ background would reach a 90\%~C.L. sensitivity to the \DBD\ half-life of more than $3\cdot 10^{26}\un{y}$, a factor 3 better than the upcoming experiment. 
Combining the light readout with an enrichment in \TEHT\ from the natural 34\% to $\sim90\%$ would push the half-life sensitivity by another factor $\sim 3$. Depending on the choice of the nuclear matrix elements, this  corresponds to an effective neutrino mass sensitivity in the range $14 -35 \un{meV}$, down into the inverted hierarchy of neutrino masses.

\section*{Acknowledgements}
The authors thank the CUORE collaboration for providing the \TEO\ crystal.
This work was supported by  the European Research Council (FP7/2007-2013) under contract  LUCIFER no. 247115
and  by the Italian Ministry of Research under the  PRIN 2010- 2011 contract no. 2010ZXAZK9.

\bibliographystyle{spphys.bst} 
\bibliography{cherenkov_bolo_ref.bbl}

\end{document}